# SOXS Control Electronics Design

G. Capasso*[a], M. Colapietro[a], S. D'Orsi[a], P. Schipani[a], M. Aliverti[b], H. Kuncarayakti[c,d], S. Scuderi[e], I. Coretti[f], S. Campana[b], R. Claudi[g], A. Baruffolo[g], S. Ben-Ami[h,i], F. Biondi[g], A. Brucalassi[j,k], R. Cosentino[l,e], F. D'Alessio[m], P. D'Avanzo[b], O. Hershko[h], M. Munari[e], A. Rubin[h], F. Vitali[m], J. Achrén[n], J. Antonio Araiza-Duran[k], I. Arcavi[o], A. Bianco[b], E. Cappellaro[g], M. Della Valle[a], O. Diner[h], D. Fantinel[g], J. Fynbo[p], A. Gal-Yam[h], M. Genoni[b], M. Hirvonen[q], J. Kotilainen[c,d], T. Kumar[d], M. Landoni[b], J. Lehti[q], G. Li Causi[m], L. Marafatto[g], S. Mattila[d], G. Pariani[b], G. Pignata[k,r], M. Rappaport[h], D. Ricci[g], M. Riva[b], B. Salasnich[g], R. Zanmar Sanchez[e], S. Smartt[s], M. Turatto[g]

[a]INAF - Osservatorio Astronomico di Capodimonte, Salita Moiariello 16, I-80131, Napoli, Italy
[b]INAF - Osservatorio Astronomico di Brera, Via Bianchi 46, I-23807 Merate (LC), Italy
[c]Finnish Centre for Astronomy with ESO (FINCA), FI-20014 University of Turku, Finland
[d]Tuorla Observatory, Department of Physics and Astronomy, FI-20014 University of Turku, Finland
[e]INAF - Osservatorio Astronomico di Catania, Via S. Sofia 78 30, I-95123 Catania, Italy
[f]INAF - Osservatorio Astronomico di Trieste, Via G.B. Tiepolo 11, I-34143 Trieste, Italy
[g]INAF - Osservatorio Astronomico di Padova, Vicolo dell'Osservatorio 5, I-35122 Padova, Italy
[h]Weizmann Institute of Science, Herzl St 234, Rehovot, 7610001, Israel
[i]Harvard Smithsonian Center for Astrophysics, Cambridge, USA
[j]ESO, Karl Schwarzschild Strasse 2, D-85748, Garching bei München, Germany
[k]Universidad Andres Bello, Avda. Republica 252, Santiago, Chile
[l]FGG-INAF, TNG, Rambla J.A. Fernández Pérez 7, E-38712 Breña Baja (TF), Spain
[m]INAF - Osservatorio Astronomico di Roma, Via Frascati 33, I-00078 Monte Porzio Catone, Italy
[n]Incident Angle Oy, Capsiankatu 4 A 29, FI-20320 Turku, Finland
[o]Tel Aviv University, Tel Aviv, Israel
[p]DARK Cosmology Center, Juliane Maries Vej 30, 2100 Copenhagen, Denmark
[q]ASRO (Aboa Space Research Oy), Tierankatu 4B, FI-20520 Turku, Finland
[r]Millennium Institute of Astrophysics (MAS), Nuncio Monseñor Sótero Sanz 100, Providencia, Santiago, Chile
[s]Astrophysics Research Centre, Queen's University Belfast, Belfast, County Antrim, BT7 1NN, UK

**ABSTRACT**

SOXS (Son Of X-Shooter) is a unique spectroscopic facility that will operate at the ESO New Technology Telescope (NTT) in La Silla from 2020 onward. The spectrograph will be able to cover simultaneously the UV-VIS and NIR bands exploiting two different arms and a Common Path feeding system. We present the design of the SOXS instrument control electronics. The electronics controls all the movements, alarms, cabinet temperatures, and electric interlocks of the instrument. We describe the main design concept. We decided to follow the ESO electronic design guidelines to minimize project time and risks and to simplify system maintenance. The design envisages Commercial Off-The-Shelf (COTS) industrial components (e.g. Beckhoff PLC and EtherCAT fieldbus modules) to obtain a modular design and to increase the overall reliability and maintainability. Preassembled industrial motorized stages are adopted allowing for high precision assembly standards and a high reliability. The electronics is kept off-board whenever possible to reduce thermal issues and instrument weight and to increase the accessibility for maintenance purpose.

The instrument project went through the Preliminary Design Review in 2017 and is currently in Final Design Phase (with FDR in July 2018). This paper outlines the status of the work and is part of a series of contributions describing the SOXS design and properties after the instrument Preliminary Design Review.

*giulio.capasso@inaf.it; phone +39 081 5575537; www.inaf.it

# 1. INTRODUCTION

SOXS (Son of X-Shooter) [1][2] will be a new spectrograph for the ESO NTT telescope able to cover the UV-VIS and NIR bands ranging from 0.35 to 2.0 µm. The task of the instrument electronics is the control of all the movements, alarms, cabinet temperatures, and electric interlocks of the instrument. Further details on specific subsystems can be found in [3-11].

This work is organized as follows: Sect. 2 presents the overall architecture; Sect. 3 presents the main PLC devoted to the system control; Sect. 4 describes all the motorized functions, the other instrument functionalities and the interconnections with the other subsystems; Sect. 5 describes the cabling; Sect. 6 describes other hardware issues (e.g. racks design).

# 2. OVERALL ELECTRONIC ARCHITECTURE

The hardware relies on Commercial Off-The-Shelf (COTS) industrial components, following the ESO design guidelines, in order to reduce costs, development time and maintenance, and to improve reliability and compatibility with the observatory standards. New instruments developed for ESO telescopes in the last few years proved the validity of the architectural solutions adopted for SOXS (e.g. [12]). The only custom components, not commercially available, are cables and signal conditioning boards where needed.

The Instrument Control Electronics (ICE) is based on one main PLC and some I/O modules connected to all subsystems. The modules are connected to the main PLC via the EtherCAT fieldbus.

The PLC offers an OPC-Unified Architecture interface on the LAN. The Instrument Software (INS) [11] installed on the Instrument Workstation (IWS) uses this protocol to send commands and read the status to/from the PLC.

Electromagnetic compatibility, safety issues, and accessibility for maintenance purpose have driven the whole design. Table 1 lists the functions to control.

Table 1. Functions to control.

| **Actuators** | | |
|---|---|---|
| Function | Sub-system | Type |
| Instrument Shutter | Common Path | Uni-stable |
| Calibration Box Selector | Common Path | Linear stage |
| Acquisition Camera Selector | Common Path | Linear stage |
| ADC Prism 1 | Common Path | Rotary stage |
| ADC Prism 2 | Common Path | Rotary stage |
| NIR Arm Focusing | Common Path | Linear stage |
| Flexure Comp. UV-VIS Arm | Common Path | Piezo Tip-Tilt |
| Flexure Comp. NIR Arm | Common Path | Piezo Tip-Tilt |
| Pinhole Exchanger | Calibration Box | Linear stage |
| Acquisition Camera Focusing | Acquisition Camera | Linear stage |
| Acquisition Camera Filter Exchanger | Acquisition Camera | Rotary stage |
| UV-VIS Slit Exchanger | UV-VIS arm | Linear stage |
| NIR Slit Exchanger | NIR Spectrograph | Linear Piezo Cryo Stage |
| **Other functions** | | |
| Function | Sub-system | Type |
| Common Path temperature sensor | Common Path | PT100 temperature sensor |
| Electronic cabinets cooling control | Electronics | Managed by the cooling controller |
| Electronic cabinets switches | Electronics | Managed by the cooling controller |
| Spectral Lamps control &diag. sensors | Calibration Box | ON/OFF-brightness adjust Fault detector and counter |
| UV-VIS Spectrograph shutter | UV-VIS Spectrograph | ESO NGC + shutter driver |
| UV-VIS/NIR detector temp. sensor | UV-VIS/NIR Spectrographs | PID integrated in temperature controller |
| Cryo-vacuum temp. sensors & control | UV-VIS/NIR Spectrographs | Independent PLC |
| Co-rotator control electronics | Co-rotator | Dedicated driver |

The PLC control system overview is shown in Figure 1.

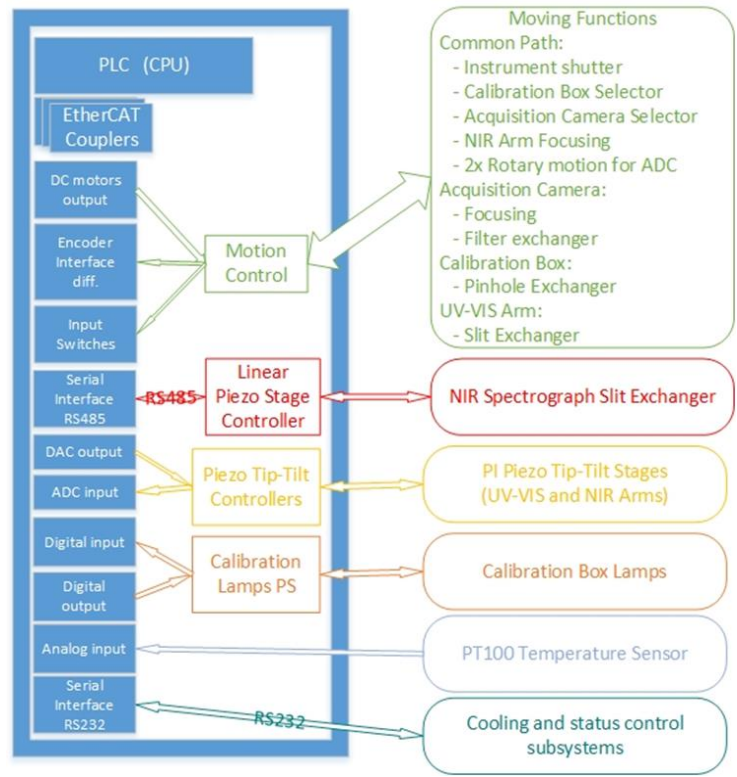

Figure 1. Control system overview.

## 3. THE PLC SYSTEM

The PLC CPU is the Beckhoff CX2030 series [13] as indicated by the ESO standard for non-high-CPU-demanding applications, coupled with the CX2100 power supply module.

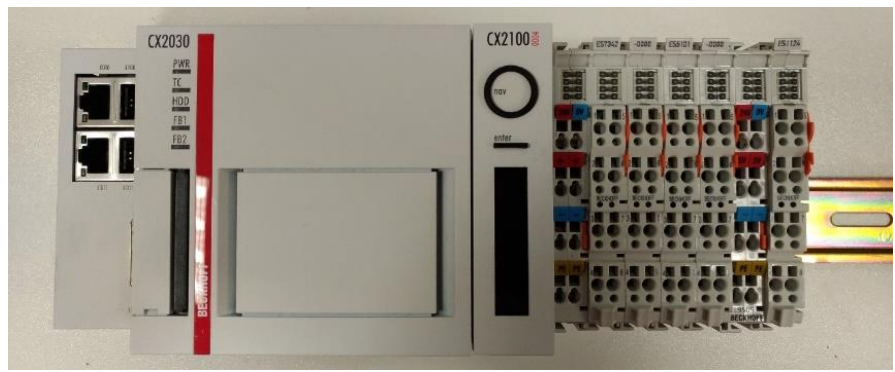

Figure 2. Beckhoff PLC and I/O modules used in the test bench.

The PLC communicates with a number of I/O modules via the EtherCAT bus.
The PLC uses the Windows operating environment with the Beckhoff TwinCAT 3 NC suite. The suite includes a series of instruments and functions for the real-time numeric control and regulation of single axis or of synchronized groups of axis. The suite includes also:
- the Point-to-point (PTP) axis positioning software;
- the OPC-UA server.

The diagram in Figure 3 shows the Beckhoff modules and their association with the controlled devices.

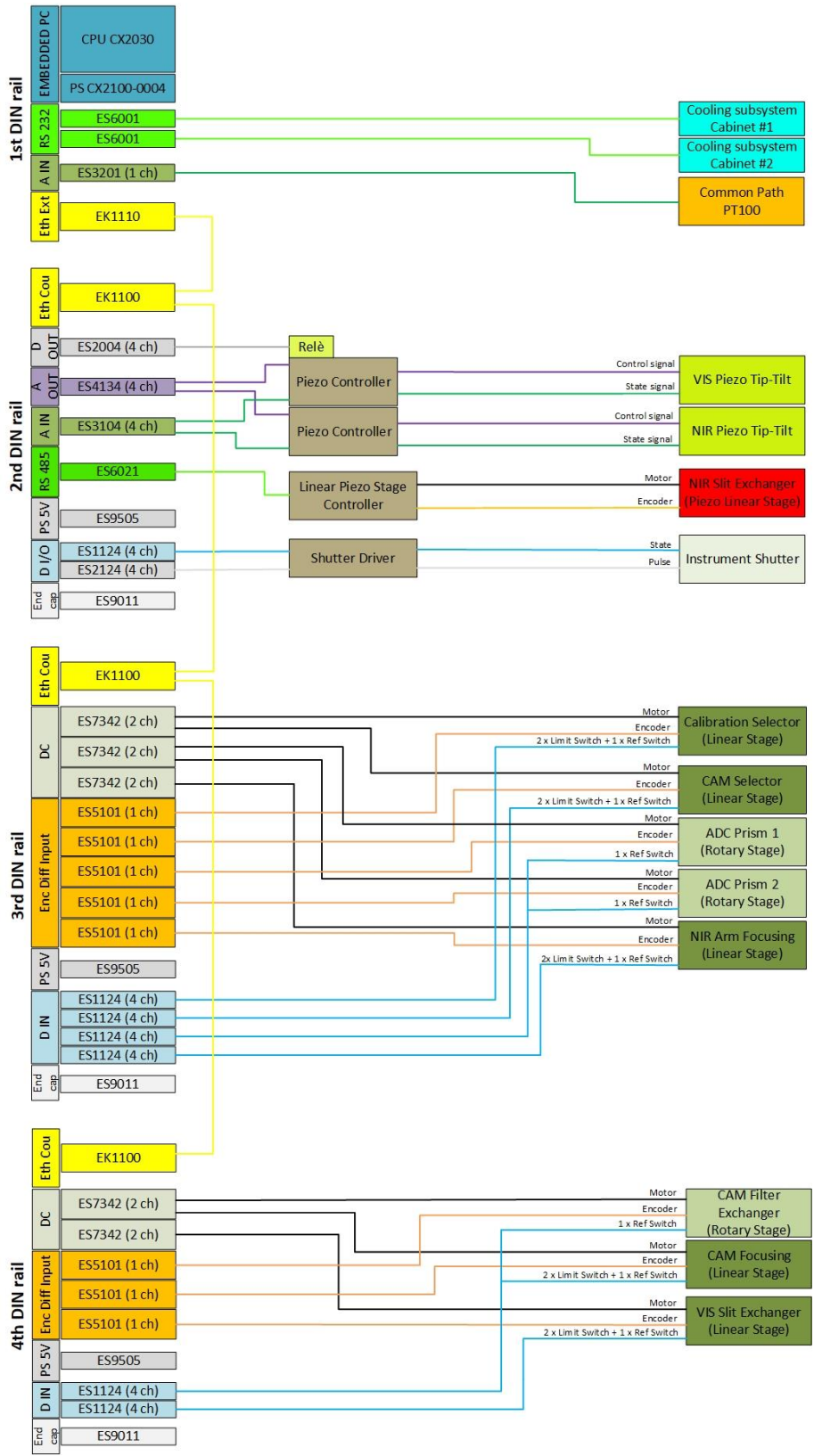

Figure 3. PLC system diagram.

## 4. CONTROLLED FUNCTIONS

### 4.1 Motorized functions control

Overall, SOXS has 13 actuators. All the motorized stages are from a single supplier (Physik Instrumente, PI) except for the linear cryogenic piezo stage for the NIR Slit Exchanger (Micronix) [4][8]. Most of the motors are DC type coupled with an incremental encoder. Table 2 lists the configuration of each motorized function.

Table 2. Detail of motorized functions.

| MOTORIZED FUNCTIONS | | | | | |
|---|---|---|---|---|---|
| Function | Actuator Type | Limit Switch | Type | Ref. Switch | Type |
| Instrument Shutter | Uni-stable | 0 | | 0 | |
| Calibration Selector | DC + Encoder | 2 | Optical, TTL | 1 | Optical, TTL |
| CAM Selector | DC + Encoder | 2 | Optical, TTL | 1 | Optical, TTL |
| ADC Prism 1 | DC + Encoder | 0 | | 1 | Mechanical |
| ADC Prism 2 | DC + Encoder | 0 | | 1 | Mechanical |
| NIR Arm Focusing | DC + Encoder | 2 | Hall, TTL | 1 | Hall, TTL |
| Flexure Comp. VIS Arm | Piezo | 0 | | 0 | |
| Flexure Comp. NIR Arm | Piezo | 0 | | 0 | |
| CBX Pinhole Exchanger | DC + Encoder | 2 | Optical, TTL | 1 | Optical, TTL |
| CAM Focusing | DC + Encoder | 2 | Hall, TTL | 1 | Hall, TTL |
| CAM Filter Exchanger | DC + Encoder | 0 | | 1 | Optical 0V,5V |
| UV-VIS Slit Exchanger | DC + Encoder | 2 | Hall, TTL | 1 | Hall, TTL |
| NIR Slit Exchanger | Piezo Stage + Encoder | 0 | | 0 | |

Each linear stage has two limit switches and one reference switch, with the exception of the linear piezo stage that has no switches. Each rotary motor stage has n.1 reference switch and the system can freely rotate.

The DC motor control is based on the PLC Beckhoff architecture and the TwinCAT NC software which acts as an interface between the hardware modules and the PLC runtime software. On the hardware side, different I/O modules are used to drive the stages to the desired positions, reading the actual speed and position values and the status of the limit switches. On the software side, the Beckhoff TwinCAT NC software environment is used to set up a fully configurable PID control loop, which controls positioning and motion velocity [13].

The DC motors are all equipped with differential encoders with EIA-485 compatible outputs and different kind of limit/reference switches:
- optical limit/reference switches;
- mechanical reference switches;
- Hall-effect limit/reference switches.

Almost all the switches are TTL and active high. The mechanical switches are n.c. (normally closed) type and are referenced to a +5V voltage.

The Beckhoff modules of the basic motion control block are:
- ES7342 - 2 channel DC motor output stage 50 VDC, 3.5 A;
- ES5101 - Incremental Encoder Interface;
- ES9505 - Power supply terminal 5 VDC, 0.5A;
- ES1124 - 4 channel digital input terminal 5 VDC.

ES7342 modules can control two DC motors and need to be configured with specific motor parameters, like nominal voltage, nominal speed, nominal and maximum current. These modules foresee a separate power supply input for the drive of the motor coil. All the motor coils are then supplied by a separate power supply, different from the one used for the logic.

## 4.2 Flexure compensation system

A flexure compensation system made by a piezo tip-tilt actuator is present in both UV-VIS and NIR arms to compensate for mechanical flexures.
The two tip-tilt stages are driven by two dedicated controllers through analog I/O interfaces.
The Beckhoff modules selected for this application are:
- ES4134 - 4 channels analog output terminal;
- ES3104 - 4 channels analog input terminal;
- ES2004 - 4 channels digital output terminal.

The two tip-tilt stage controllers are driven by means of an analog loop: DACs (ES4134) give reference positions to the controllers and ADCs (ES3104) read the actuator positions.
On the other side, the controllers interact with the actuators, again through analog signals: a control signal sends the reference positions to the piezo actuators, while a state signal returns the actual positions back.
A supplementary relay, driven by a digital output (ES2004), allows for the subsystem shutdown in order to reduce usage and power consumption. For the two tip-tilt stages, the general scheme in Figure 4 is adopted.

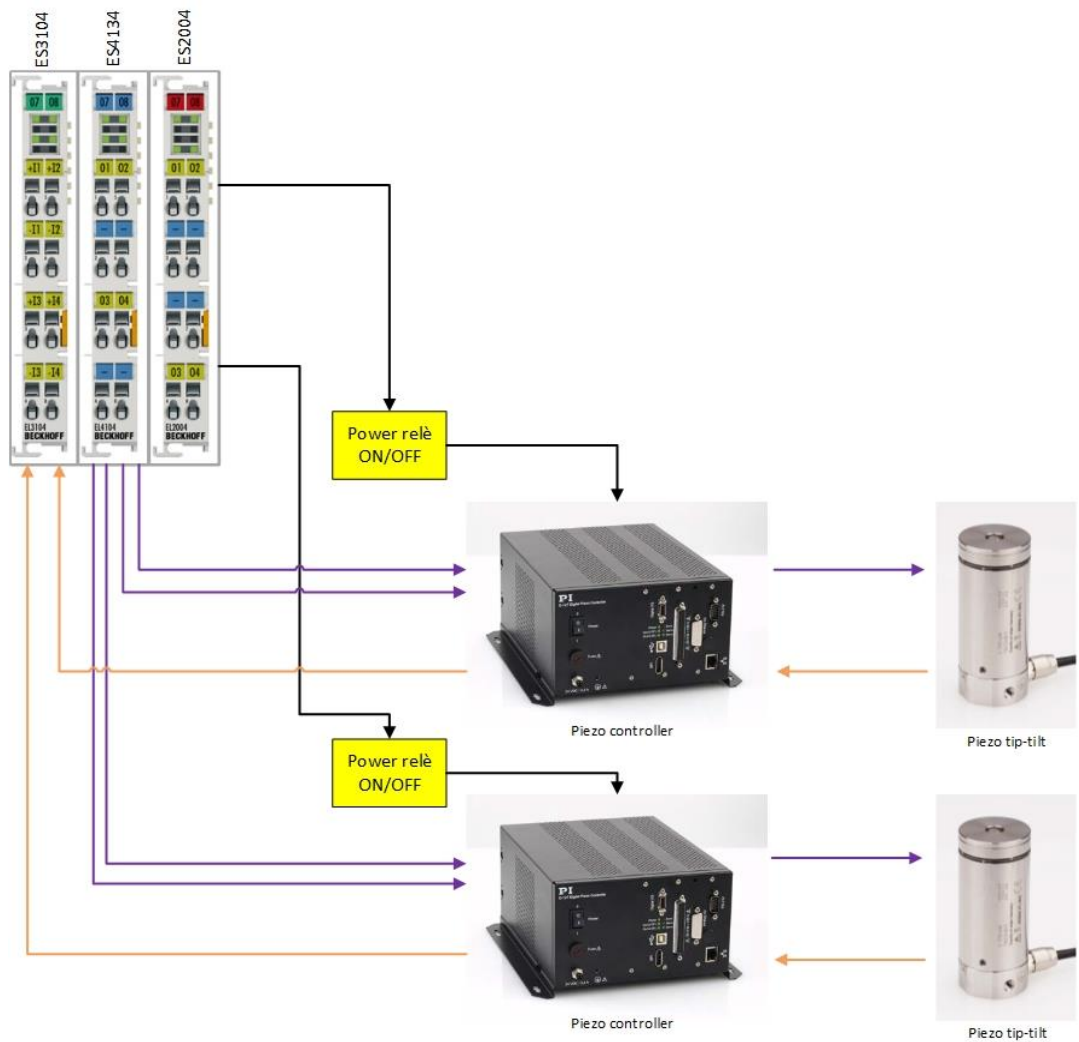

Figure 4. Piezo tip-tilt control architecture.

## 4.3 Co-rotator control electronic

The SOXS instrument will be mounted on the Nasmyth interface of the NTT telescope and will rotate following the parallactic angle. Therefore, an appropriate co-rotator structure is needed to drive cables and pipes [8].
The co-rotator control electronics is based on a servo motor driven by a dedicated controller.
Two linear potentiometers provide feedback signals for the differential position between the instrument and the co-rotator which are appropriately combined by a custom PCB adapter board and then sent to the drive as speed reference.
A mechanical switch is foreseen for safety reasons when the misalignment between the two systems exceeds a configurable threshold. It cuts off the power if the control loop fails. The general control diagram is shown in Figure 5. The co-rotator control electronics will be hosted in a small dedicated box, placed close to the motor.

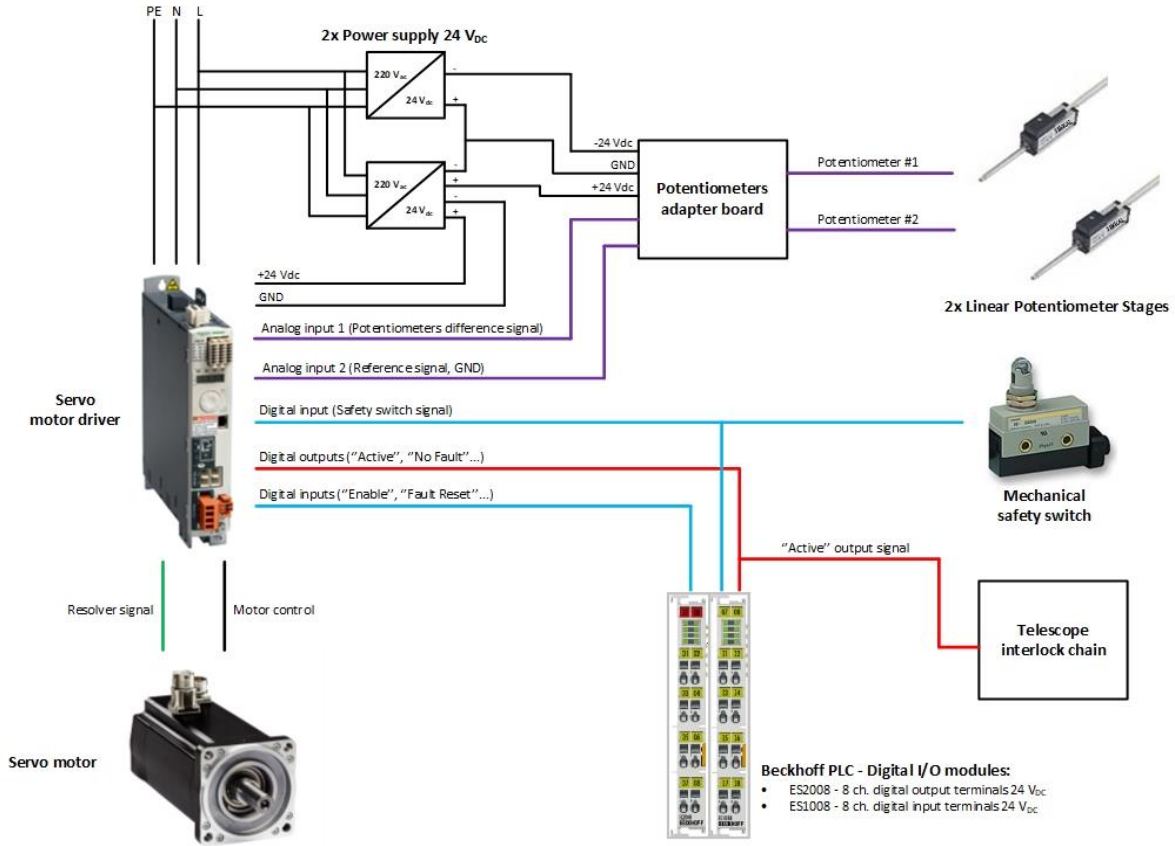

Figure 5. General co-rotator motor control diagram.

## 4.4 Cryo-vacuum

A Siemens S7-1500 PLC will control the Cryogenic and Vacuum subsystem. This PLC will control all the cryo-vacuum electronics (gauges, valves, turbo-molecular pump, pre-vacuum pump and cryo-cooler compressor) and will meet safety standards. The cryo-vacuum system will also control the temperature inside the NIR Spectrograph through heaters and sensors.
Two Lakeshore 336 temperature controllers handle the temperature of the two detectors (UV-VIS and NIR) using PID control loops.
The cryo-vacuum control subsystem will be integrated in one of the two electronic racks installed on the Nasmyth platform. All communications with the instrument software will occur through the LAN via OPC-UA.

**4.5 Calibration lamps**

The Calibration Unit (CBX) includes several calibration and flat-fielding lamps and one motorized stage. The control system has to switch ON/OFF the lamps and control the stage. Lamp fault detectors, a safety interlock signal, and a current sensing circuit for monitoring the state of the lamps complete this subsystem.

The CBX electronic subsystem follows the same general architecture shown previously. Beckhoff I/O modules are linked via EtherCAT bus to the main PLC.

The CBX control system is placed in a separate sub-rack hosted in one of the electronic cabinets. Such modular design simplifies Assembly Integration and Tests at different consortium premises.

**4.6 Detectors**

The electronics for the two UV-VIS and NIR detectors is based on the New Generation Controller (NGC) supplied by ESO. We take care of cable routing and cable length to protect detector signals from EMC interferences. The NGCs will be mounted on the instrument, near the detectors; the two power supplies will be installed within the electronic racks.
The UV-VIS NGC controls the UV-VIS spectrograph shutter as well.

## 5. CABLING

The motor stage cabling handles:
- DC motor power supply;
- differential encoder signals;
- limit and/or reference switches signals;
- encoder & switches power supply.

In our cabling design we needed to consider the following issues.
- To reduce the instrument weight, most of electronics has been moved off board, into racks on the Nasmyth platform. So the length of most of the cables is approx. 10 meters and we need to take care of possible noise and voltage drops.
- To manage the large number of cables a cable wrap is used between the instrument and the racks. This implies the use of highly flexible cables. Also, cables must use auto extinguish insulator for safety purpose.
- COTS stages come with their own cables and controllers; cables use a unique shield to cover both power and signal conductors, producing high cross-talk interference when used with our setup.
- Most of the motorized stages use TTL for limit and reference switches; these low voltage signals suffer for EMC interferences.

During our lab tests, COTS cables showed noise due to an interference coming from the motor power supply, that could be a critical issue. Instead, encoder signals showed immunity to noise, thanks to their differential levels. This led us to a different cabling strategy. We decided to design and manufacture custom cables with higher immunity to noise.

To simplify maintenance, we also decided to keep the same connector and cabling schema whenever possible.
Cables and connectors will have colored labels according to a specific policy.
The cable labels are of the type: **SSS-FFFF-YY**
- Each subsystem has its own prefix SSS (e.g. CPT for the Common Path);
- The elements of each sub-system have their own suffix -FFFF (e.g. CALS for the Calibration Selector);
- Finally, a number (-YY) identifies uniquely the cable.

The connector IDs have the following code: **SSS-FFFF/XX**
where the /XX is a combination (letter + number) that identifies uniquely the connector (e.g. "A1", "B2", etc.).

## 6. CONTROL RACKS

Almost all the SOXS Control Electronics is hosted in two standard 19″ cabinets. The exceptions are the two ESO NGCs that are mounted on the instrument and the co-rotator electronics that is hosted in a dedicated box.
The cabinets are divided in sub-racks (see Figure 6) in order to simplify the maintenance and the test phase of the sub-systems, during which some parts of the control electronics will need to travel between different consortium institutes.

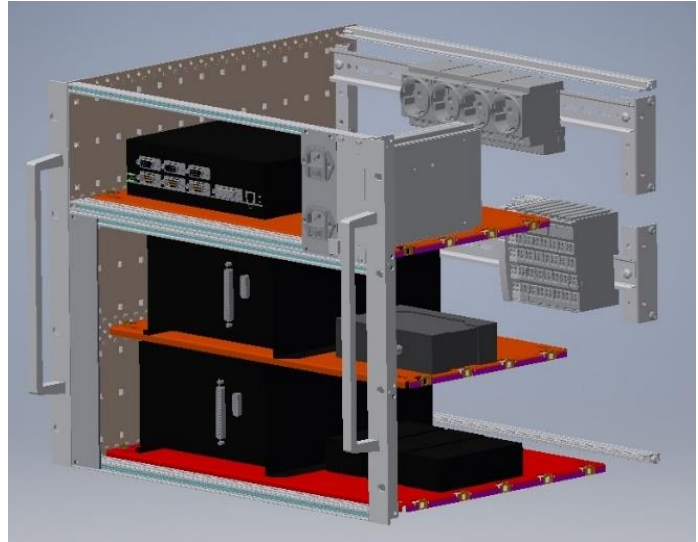
Figure 6. Example of subrack organization.

A first cabinet hosts:
- a 9U sub-rack containing all the dedicated controllers (shutter, tip-tilt and linear piezo stages);
- a 9U sub-rack containing the Beckhoff CPU, a control panel and all the modules to control the other motorized stages;
- a 9U sub-rack containing all the Calibration Unit electronics (lamp power supplies and the Beckhoff modules);
- the two NGC Linux machines for the UV-VIS and NIR detectors;
- a network switch;
- the cabinet thermal control unit.

A second cabinet hosts:
- the two NGC power supply;
- a sub-rack containing all the Cryo-Vacuum control electronics;
- a network switch;
- the cabinet thermal control unit.

Figure 7 shows the two racks design. Both cabinets are equipped with their own integrated cooling system. Cabinet temperature, door switches and the flow of the coolant liquid are controlled through a dedicated ESO Thermal Control Unit. Additional electronics include:
- an Ethernet switch that will connect all subsystems using a private IP addressing scheme.
- the power supply, both normal and UPS; each line is filtered and protected against over-voltage, over-current, and over-temperature.

## 7. CONCLUSION

The ICE is based on a COTS PLC from Beckhoff and controls all instrument movements, functionalities and interlocks. There are a number of I/O modules, connected via an EtherCAT network. The electronic hardware is mainly hosted in two racks on the Nasmyth platform, with few parts onboard. The electronics and cables design has to comply with the necessary EMC requirements.
The PLC is controlled by the Instrument Software through the OPC-UA interface.

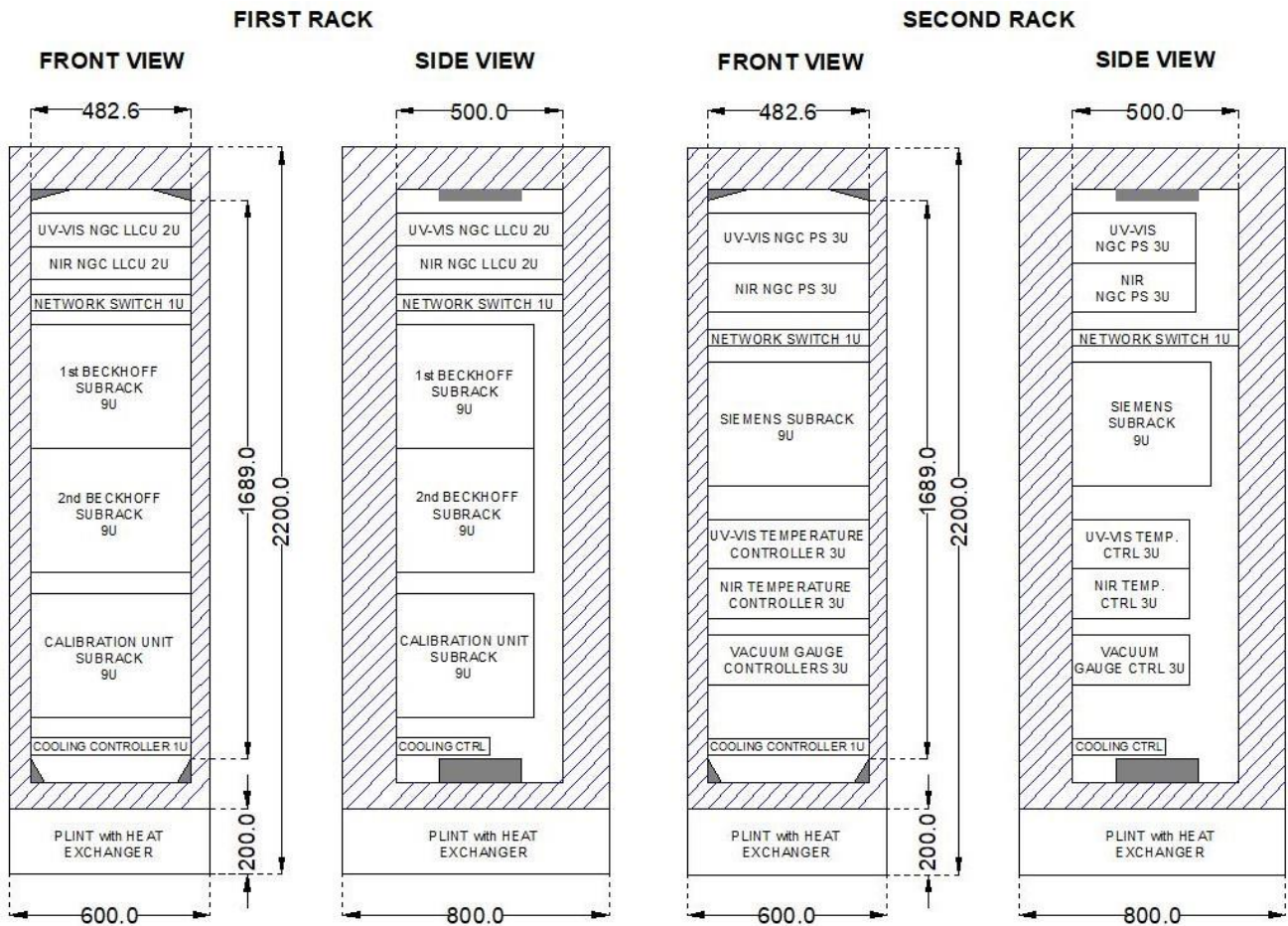

Figure 7. Electronics racks organization.